\documentclass[10pt,conference,twocolumn,letterpaper]{IEEEtran}
\IEEEoverridecommandlockouts

\usepackage{amsmath}
\usepackage{amsfonts}
\usepackage{amssymb}
\usepackage{amsthm}
\usepackage{tabularx}
\usepackage{mathrsfs} 
\usepackage{soul}
\usepackage{graphicx}

\usepackage[draft,bookmarks=false]{hyperref}  % draft disables link creation
\usepackage{cleveref}

\usepackage{hyperref}

\usepackage{stfloats}
\usepackage{float}
\usepackage{graphicx}
\usepackage{cite}
\usepackage{xcolor}
\usepackage{subfigure}

\usepackage{cleveref}

\usepackage{threeparttable}

\usepackage{algorithm}
\usepackage{algpseudocode}

\usepackage{pifont}

\usepackage{booktabs}

\usepackage{multirow}

\usepackage{rotating}

\usepackage{makecell}

\usepackage{array}

\makeatletter
\def\ps@headings{}                % ensure IEEEtran doesn't reserve header space
\makeatother

\def\BibTeX{{\rm B\kern-.05em{\sc i\kern-.025em b}\kern-.08em
    T\kern-.1667em\lower.7ex\hbox{E}\kern-.125emX}}
\usepackage{balance}

\begin{document}
\title{Fluid Antenna-Enabled Backscatter Communication under Imperfect Channel Observations}
\author{
\IEEEauthorblockN{
Masoud Kaveh\IEEEauthorrefmark{2},
Farshad Rostami Ghadi\IEEEauthorrefmark{4},
Riku J\"antti\IEEEauthorrefmark{2},
Kai-Kit Wong\IEEEauthorrefmark{3},
F. Javier Lopez-Martinez\IEEEauthorrefmark{4}
}
\IEEEauthorblockA{\IEEEauthorrefmark{2}Department of Information and Communications Engineering, Aalto University, Espoo, Finland}
\IEEEauthorblockA{\IEEEauthorrefmark{4}Department of Signal Theory, Networking and Communications, University of Granada, Granada, Spain. }
\IEEEauthorblockA{\IEEEauthorrefmark{3}Department of Electronic and Electrical Engineering, University College London, London, U.K.}

}

%\markboth {How to Use the IEEEtran \LaTeX \ Templates}
%{Journal of \LaTeX\ Class Files,~Vol.~xx, No.~xx, September~20xx}%

\maketitle

\begin{abstract}
Ambient backscatter communication (AmBC) enables ultra-low-power connectivity by allowing passive backscatter devices (BDs) to convey information through reflection of ambient signals. However, the cascaded AmBC channel suffers from severe double path loss and multiplicative fading, while accurate channel state information (CSI) acquisition is highly challenging due to the weak backscattered signal and the resource-limited nature of BDs. To address these challenges, this paper considers an AmBC system in which the reader is equipped with a pixel-based fluid antenna system (FAS). By dynamically selecting one antenna position from a dense set of pixels within a compact aperture, the FAS-enabled reader exploits spatial diversity through measurement-driven port selection, without requiring explicit CSI acquisition or multiple RF chains. The intrinsic rate-energy tradeoff at the BD is also incorporated by jointly optimizing the backscatter modulation coefficient under an energy harvesting (EH) neutrality constraint. To efficiently solve this problem, a particle swarm optimization (PSO)-based framework is developed to jointly determine the FAS port selection and modulation coefficient on an optimize-then-average (OTA) basis. Simulation results show that the proposed scheme improves the achievable rate compared with fixed antenna readers, with gains preserved under imperfect observations, stringent EH constraints, and different pixel spacings.
\end{abstract}


\section{Introduction}
\IEEEPARstart{A}{mbient} backscatter communication (AmBC) has emerged as a key enabling technology for ultra-low-power and battery-less connectivity in large-scale Internet-of-Things (IoT) networks \cite{Jamshed_AmBC_NTN_2025,Liao_AmBC_LTE_2025}. By modulating and reflecting incident radio-frequency (RF) signals instead of generating new waveforms, backscatter devices (BDs) can operate with minimal energy consumption and hardware complexity \cite{Kaveh2025:VPD-PLA}. Despite these advantages, AmBC performance is fundamentally limited by the cascaded channel formed by source-to-BD and BD-to-reader links. The resulting double path loss and multiplicative fading cause severe signal attenuation, which significantly constrains AmBC coverage and reliability \cite{Gu_BC_Survey_2025,Kaveh2024:SecureRIS}. In addition, acquiring accurate instantaneous channel state information (CSI) is particularly challenging due to the extremely weak backscattered signal and the inability of passive BDs to support explicit channel training \cite{Guo_MIMO_AmBC_2018,Ma_Blind_CE_AmBC_2018}.

Recent studies have explored fluid antenna systems (FAS) \cite{Wong2021:FAS,Fluid-survey}
as an effective means of mitigating channel impairments in backscatter and energy-constrained wireless systems \cite{Ghadi_FAS_AmBC_2024,Ghadi_FAS_NOMA_WiMob_2024,Ghadi_FAS_PLS_2025}. Early analytical works show that antenna position flexibility at the reader can significantly improve outage performance and reliability in backscatter communications (BC) under spatially correlated fading, confirming the diversity gains of FAS over conventional fixed-antenna readers \cite{Ghadi_FAS_AmBC_2024}. Related studies have extended FAS concepts to wireless powered systems, demonstrating notable improvements in outage probability, secrecy performance, and achievable capacity when a single FAS port is optimally selected \cite{Ghadi_FAS_NOMA_WiMob_2024,Ghadi_FAS_PLS_2025}.

Despite the promising performance gains reported in \cite{Ghadi_FAS_AmBC_2024,Ghadi_FAS_NOMA_WiMob_2024,Ghadi_FAS_PLS_2025}, existing FAS-aided backscatter studies predominantly rely on idealized assumptions, such as analytically tractable fading models and perfect or statistical channel knowledge. In particular, the effect of channel gain observation uncertainty, an inherent characteristic of AmBC arising from the extremely weak backscattered signal and the absence of explicit channel training at passive BDs, has received little attention. Furthermore, prior works typically decouple communication design from the energy harvesting (EH) requirements of the BD by assuming fixed or unconstrained reflection coefficients. In practical AmBC deployments, however, the BD must sustain its operation through harvested energy while simultaneously enabling reliable information transfer, giving rise to an intrinsic rate–energy tradeoff that directly impacts system performance \cite{Ghosh_EH_NOMA_AmBC_2025,Kaveh_Secrecy_BC_Sensors_2023}. These considerations expose a clear gap between existing analytical FAS performance characterizations and the development of robust, implementable antenna selection strategies suited to realistic AmBC conditions.

Motivated by this gap, this paper investigates an AmBC system in which the reader is equipped with a pixel-based FAS to enhance the cascaded backscatter channel under practical operating constraints. Instead of assuming perfect CSI, antenna port selection is performed based solely on noisy observations of the cascaded channel gains. To explicitly account for energy sustainability at the BD, we formulate a reader-centric optimization problem that maximizes the achievable rate by jointly optimizing the FAS port selection and the backscatter modulation coefficient under an EH constraint. Owing to the nonconvex nature of the resulting problem, a particle swarm optimization (PSO)-based framework is developed that operates directly on observed channel gains. Simulation results confirm that the proposed FAS-enabled reader consistently outperforms traditional fixed-antenna benchmarks across a wide range of operating conditions.

%%%%%%%%%%%%%%%%%%%%%%%%%%%%
\begin{figure}[t]
    \centering    \includegraphics[width=0.44\textwidth]{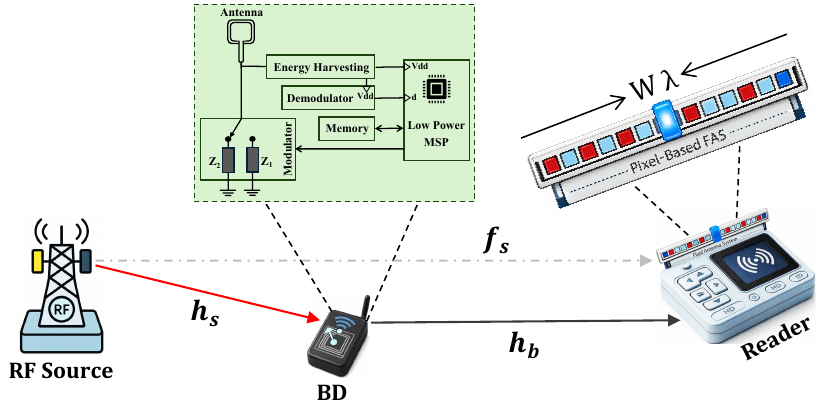}
    \caption{The considered AmBC system with pixel-based FAS at the reader.}
    \label{fig:sysmodel}
\end{figure}
%%%%%%%%%%%%%%%%%%%%%%%%%%%%
\section{System Model}
\label{sec:sysmodel}

We consider an AmBC setup consisting of an RF source (illuminator), a single-antenna passive BD, and a reader equipped with a pixel-based FAS, as illustrated in Fig.~\ref{fig:sysmodel}.
The RF source continuously transmits an unmodulated carrier (or a known waveform) with transmit power $P_\mathrm{s}$.
The BD harvests energy from the incident RF signal and conveys information by switching its load impedance between two states, corresponding to binary backscatter modulation \cite{Barbot_Backscatter_Receiver_2026,Kaveh_RIS_T2T_MeditCom_2025}.
The reader employs the FAS to activate one antenna port (pixel) at a time from a set of $K$ preset ports distributed over a linear aperture of length $W\lambda$, where $\lambda$ denotes the carrier wavelength and $W$ is the normalized FAS size.
The $K$ FAS ports are uniformly spaced along the aperture with sub-wavelength inter-port distance $d$ \cite{Ghadi_Copula_FAS_2023}.

%\subsection{Channel Model}
Let $h_\mathrm{s}\in\mathbb{C}$ denote the forward channel from the RF source to the BD, and $h_{\mathrm{b},k}\in\mathbb{C}$ denote the backscatter channel from the BD to the $k$-th FAS port at the reader, where $k\in\{1,\dots,K\}$.
The corresponding channel power gains are defined as
\begin{equation}
g_\mathrm{s} \triangleq |h_\mathrm{s}|^2,\qquad
g_{\mathrm{b},k} \triangleq |h_{\mathrm{b},k}|^2 .
\end{equation}
Due to the finite aperture of the FAS, the backscatter channels $\{h_{\mathrm{b},k}\}$ are spatially correlated.
A commonly adopted correlation model is the Jakes' correlation across antenna ports, given by
\begin{equation}
\mathbf{R} = \big[\rho_{k,\ell}\big],\qquad
\rho_{k,\ell} = J_0\!\left( 2\pi \frac{|k-\ell|}{K-1} W \right),
\label{eq:jakes}
\end{equation}
where $J_0(\cdot)$ denotes the zero-order Bessel function of the first kind, and $\rho_{k,\ell}$ represents the spatial correlation coefficient between the $k$-th and $\ell$-th FAS ports.

%\subsection{Backscatter Modulation and Cascaded Channel}
Then, the RF signal received at the BD is expressed as
\begin{equation}
r_\mathrm{BD} = \sqrt{P_\mathrm{s}}\, h_\mathrm{s}\, x + f_\mathrm{s} + n_\mathrm{BD},
\end{equation}
where $x$ is the unit-power transmitted waveform, i.e., $\mathbb{E}[|x|^2]=1$, $f_\mathrm{s}$ is the direct link from RF source to the reader, and $n_\mathrm{BD}$ denotes the receiver noise at the BD, which is typically negligible compared to the incident carrier for passive operation \cite{Kaveh2024:SecureRIS,Liu_Tag_Selection_PLS_2021}. $f_\mathrm{s}$ is assumed to be suppressed via well-established cancellation techniques in backscatter receivers \cite{Guo2022}.
The BD modulates information by switching its load impedance, thereby changing its reflection coefficient.
Let $\Gamma_i$ denote the reflection coefficient associated with impedance state $Z_i$, $i\in\{0,1\}$ \cite{Zhang_Phy_CRA_AmBC_2026}.
For binary signaling, the reflection coefficient selected by the BD is
\begin{equation}
\Gamma(b) =
\begin{cases}
\Gamma_0, & b=0,\\
\Gamma_1, & b=1,
\end{cases}
\end{equation}
where $b\in\{0,1\}$ denotes the information bit. In this work, we consider ON-OFF keying (OOK) backscatter modulation \cite{Guo_LoRa_AmBC_2021}, where the \emph{OFF} state ($b=0$) corresponds to a matched load with negligible reflection, and the \emph{ON} state ($b=1$) corresponds to a mismatched load that reflects a portion of the incident RF signal.
The backscatter modulation coefficient is defined as
\begin{equation}
\alpha(b) \triangleq \sqrt{\eta}\,\Gamma(b),
\end{equation}
where $\eta\in(0,1]$ accounts for the reflection efficiency of the BD, capturing losses due to impedance mismatch and non-ideal reradiation.

Consequently, the received baseband signal at the $k$-th FAS port is modeled as
\begin{equation}
y_k = \sqrt{P_\mathrm{s}}\, h_\mathrm{s}\, h_{\mathrm{b},k}\, \alpha(b)\, x + z_k,
\label{eq:yk}
\end{equation}
where $z_k\sim\mathcal{CN}(0,\sigma^2)$ is additive white Gaussian noise (AWGN) at the reader.
Accordingly, the cascaded end-to-end channel observed at the $k$-th port is
\begin{equation}
h_{\mathrm{c},k} \triangleq h_\mathrm{s}\, h_{\mathrm{b},k},
\qquad
g_{\mathrm{c},k} \triangleq |h_{\mathrm{c},k}|^2 = g_\mathrm{s}\, g_{\mathrm{b},k}.
\label{eq:cascaded}
\end{equation}

%\subsection{Measurement-Driven FAS Port Selection}
The pixel-based FAS at the reader is equipped with a single RF chain and can activate only one antenna port at a time.
Due to the passive nature of the BD and the extremely weak backscattered signal, acquiring accurate instantaneous channel state information (CSI) is highly challenging in AmBC systems.
Therefore, rather than assuming perfect CSI or performing explicit channel estimation, the reader adopts a measurement-driven port selection strategy based on noisy observations of the port quality, which are obtained through short-term measurements of received signal-to-noise ratio (SNR).
Specifically, the observed cascaded gain at the $k$-th port is modeled as
\begin{equation}
\tilde g_{\mathrm{c},k} = g_{\mathrm{c},k}\big(1+\delta_k\big),
\label{g-noisee}
\end{equation}
where $\delta_k\sim\mathcal{N}(0,\sigma_\delta^2)$ models the relative uncertainty in the observed port quality due to noisy power/SNR measurements of the cascaded channel \cite{Chu_RSS_Multiplicative_2022}. This model reflects a measurement-driven observation process rather than classical channel estimation, and captures imperfect channel knowledge at the reader without assuming access to instantaneous CSI.
Then, the reader selects the FAS port according to
\begin{equation}
k^\star = \arg\max_{k\in\{1,\dots,K\}} \; \tilde g_{\mathrm{c},k},
\label{eq:select}
\end{equation}
which yields the effective cascaded channel gain
\begin{equation}
g_\mathrm{FAS} \triangleq g_{\mathrm{c},k^\star}.
\label{eq:gFAS}
\end{equation}
The post-selection received signal at the reader is given by
\begin{equation}
y = \sqrt{P_\mathrm{s}}\, h_{\mathrm{c},k^\star}\,\alpha(b)\,x + z,
\label{eq:ysel}
\end{equation}
where $z\sim\mathcal{CN}(0,\sigma^2)$ denotes AWGN.

%\subsection{Achievable Rate and Energy Harvesting}
Conditioned on the selected FAS port and the backscatter coefficient, the instantaneous SNR at the reader is
\begin{equation}
\gamma = \frac{P_\mathrm{s}\, |\alpha(b)|^2\, g_\mathrm{FAS}}{\sigma^2}.
\label{eq:snr}
\end{equation}
Accordingly, the achievable backscatter rate is expressed as
\begin{equation}
R = \log_2\!\left(1+\gamma\right),
\label{eq:rate}
\end{equation}
which serves as an upper-bound performance metric commonly adopted to evaluate relative gains under different antenna and observation configurations.

In addition to information transfer, the BD harvests energy from the incident RF signal.
The average RF power impinging on the BD antenna is proportional to $P_\mathrm{s} g_\mathrm{s}$.
Since AmBC relies on reflecting a portion of the incident signal, the harvested and backscattered powers originate from the same RF source and therefore compete for the available energy at the BD.
To capture this trade-off, an EH model is adopted \cite{RIS-AmBC-Rezaei}, where the harvested DC power is given by\footnote{In this work, a linear EH model is adopted for simplicity, whereas practical non-linear EH characteristics have been investigated in \cite{Alevizos2018RFEH}.}
\begin{equation}
P_\mathrm{EH} = \xi \big(1-|\alpha(b)|^2\big) P_\mathrm{s} g_\mathrm{s},
\label{eq:eh}
\end{equation}
with $\xi\in(0,1]$ denoting the RF-to-DC conversion efficiency.
To ensure energy-neutral operation of the BD, the harvested power must satisfy
\begin{equation}
P_\mathrm{EH} \ge P_\mathrm{c},
\label{eq:eneutral}
\end{equation}
where $P_\mathrm{c}$ denotes the circuit power consumption associated with sensing, logic, and impedance switching.
%%%%%%%%%%%%%%%%%%%%%%%%%%%%%%%
%|\alpha| is optimized as a continuous variable to characterize the optimal operating point; in practice, the BD selects the nearest feasible impedance state.
%%%%%%%%%%%%%%%%%%%%%%%%%%%%%%% 

\section{Problem Formulation}
\label{sec:problem}

In this section, we formulate the joint optimization problem for the considered pixel-based FAS-enabled AmBC system.
Our objective is to maximize the achievable backscatter rate by jointly optimizing the backscatter modulation coefficient at the BD and the antenna port selection at the FAS-equipped reader, while accounting for imperfect CSI and ensuring energy-neutral operation of the BD.

The first optimization variable is the non-zero backscatter modulation level $a_1 \triangleq |\alpha(1)|^2$, corresponding to the \emph{ON} reflection state of the BD, where $a_0=|\alpha(0)|^2=0$, and the second is the selected FAS port index $k \in \{1,\dots,K\}$ at the reader.
For practical backscatter hardware, the modulation level $a_1$ is constrained to a feasible interval \cite{Xu_Nonlinear_EH_2017}
\begin{equation}
a_{\min} \le a_1 \le a_{\max} \le 1,
\label{eq:a_bounds}
\end{equation}
where $a_{\min}$ ensures a detectable backscattered signal and $a_{\max}$ is determined by circuit and impedance design limitations.

As established earlier, due to imperfect CSI, the reader relies on noisy observations of the cascaded channel gains.
Let $\tilde g_{\mathrm{c},k}$ denote the observed cascaded gain at port $k$ (cf. \eqref{g-noisee}), and denote by $g_\mathrm{FAS}$ the effective (true) cascaded gain of the selected port.
For a given pair $(k,a_1)$, the instantaneous rate is written as $R(k,a_1)$, where the dependence on imperfect CSI is captured through the observation-driven decision on $k$.
Assuming binary backscatter modulation with $\Pr(b=1)=p_1$ and $\Pr(b=0)=1-p_1$, the instantaneous SNR follows directly from \eqref{eq:snr}, and thus $R(k,a_1)$ serves as the objective function to be maximized.

On the other hand, the BD must satisfy an energy sustainability constraint to maintain passive operation.
Under the adopted linear EH model in \eqref{eq:eh}, we impose an EH margin constraint to ensure robust operation under channel fluctuations, given by
\begin{equation}
\xi (1 - p_1 a_1) P_s g_s \ge \mu P_c
\label{eq:eh_constraint}
\end{equation}
where $\mu>1$ is a prescribed EH margin factor.

To be consistent with practical operation under imperfect CSI, we adopt an optimize-then-average (OTA) performance metric.
Specifically, for each channel realization and observation-noise sample, the reader selects a port based on the observed gains and the design variables, and the achieved rate is then averaged across realizations.
Accordingly, the joint OTA optimization problem is formulated as
\begin{subequations}
\label{prob:main}
\begin{align}
\max_{a_1,k} \quad & \mathbb{E}[R(a_1,k^\star)] \label{prob:obj} \\
\text{s.t.}\quad & \xi (1-p_1 a_1) P_s g_s \ge \mu P_c, \label{prob:eh} \\
& a_{\min} \le a_1 \le a_{\max}, \label{prob:a1} \\
& k \in \{1,\dots,K\}. \label{prob:k}
\end{align}
\end{subequations}
Problem~\eqref{prob:main} is non-convex and involves mixed discrete-continuous optimization due to the discrete FAS port index and the coupled dependence of the achievable rate and harvested energy on the backscatter modulation coefficient.
While the reflection coefficient can be optimized in closed form under EH constraints for a selected port, the port-selection variable renders the problem nonconvex and combinatorial. To efficiently handle this structure under imperfect CSI, we adopt a PSO-based solution to efficiently solve this problem.
To that aim, each particle $i$ maintains a position vector
\begin{equation}
\mathbf{x}_i = [u_i, a_{1,i}],
\end{equation}
where $u_i \in [1,K]$ is a continuous surrogate variable that is decoded to the discrete FAS port index as
$k_i=\mathrm{clip}(\lfloor u_i \rceil,1,K)$, and $a_{1,i}\in[a_{\min},a_{\max}]$ denotes the candidate ON-level backscatter modulation coefficient.
This encoding enables joint optimization of the reader-side port selection and the BD-side reflection strength within a unified search space.

At iteration $t$, the velocity and position of particle $i$ are updated according to
\begin{equation}
\mathbf{v}_i^{(t+1)} =
\omega \mathbf{v}_i^{(t)}
+ c_1 r_1 \big(\mathbf{x}_{i,\mathrm{best}}-\mathbf{x}_i^{(t)}\big)
+ c_2 r_2 \big(\mathbf{x}_{\mathrm{best}}-\mathbf{x}_i^{(t)}\big),
\label{eq:pso_v_final}
\end{equation}
\begin{equation}
\mathbf{x}_i^{(t+1)} =
\Pi_{\mathcal{X}}\!\left(\mathbf{x}_i^{(t)} + \mathbf{v}_i^{(t+1)}\right),
\label{eq:pso_x_final}
\end{equation}
where $\omega$ is the inertia weight, $c_1$ and $c_2$ are the cognitive and social learning coefficients, and $r_1,r_2\sim\mathcal{U}(0,1)$.
The projection operator $\Pi_{\mathcal{X}}(\cdot)$ enforces feasibility by ensuring
$u_i\in[1,K]$, $a_{1,i}\in[a_{\min},a_{\max}]$, and satisfaction of the energy-neutrality constraint in \eqref{prob:eh}.

For each particle realization, the decoded port index $k_i$ determines the FAS port selected at the reader based on the noisy cascaded gain observations.
The instantaneous backscatter rate corresponding to $(k_i,a_{1,i})$ is then evaluated and used as the particle fitness.
Personal and global best solutions are updated accordingly.
After convergence, the PSO yields the optimized solution $(k^\star,a_1^\star)$, whose performance is assessed using the OTA rate metric.
Algorithm~\ref{alg:pso-fas-ambc} illustrates the PSO-driven optimization process in this paper.
The computational complexity of the proposed PSO framework scales approximately as $\mathcal{O}(N N_p T)$, where $N$, $N_p$, and $T$ denote the number of channel realizations, particles, and iterations, respectively.

\begin{algorithm}[t]
\caption{PSO for Pixel-Based FAS-aided AmBC }
\label{alg:pso-fas-ambc}
\small
\begin{algorithmic}[1]
\Require $K$, $N$, $N_p$, $T$, $(\omega,c_1,c_2)$, $(P_\mathrm{s},\sigma^2,\xi,P_\mathrm{c},p_1,\epsilon)$, $\sigma_\delta^2$, $(a_{\min},a_{\max})$.
\Ensure Best solution $(k^\star,a_1^\star)$ and $\bar R_{\mathrm{OTA}}^\star$.
\For{$n=1$ to $N$}
    \State Generate one channel realization and observation noise; form $\{\tilde g_{\mathrm{c},k}\}_{k=1}^K$.
    \State Initialize particles $\mathbf{x}_i=[u_i,a_{1,i}]$ with $u_i\in[1,K]$, $a_{1,i}\in[a_{\min},a_{\max}]$.
    \For{$t=1$ to $T$}
        \For{each particle $i$}
            \State Decode $k_i \leftarrow \mathrm{clip}(\lfloor u_i\rceil,1,K)$ and set candidate $a_{1,i}$.
            \State \textit{If} \eqref{prob:eh} is violated, discard particle (infeasible); 
            \State \textit{Otherwise} compute instantaneous rate as fitness.
            \State Update personal and global bests.
        \EndFor
        \State Update particle velocities and positions;
        \State Project onto feasible ranges.
    \EndFor
    \State Record the optimized $R^{\star(n)}$ with $(k^\star,a_1^\star)$ for realization $n$.
\EndFor
\State Compute $\bar R_{\mathrm{OTA}}^\star = \frac{1}{N}\sum_{n=1}^N R^{\star(n)}$.
\State \Return $(k^\star,a_1^\star)$ and $\bar R_{\mathrm{OTA}}^\star$.
\end{algorithmic}
\end{algorithm}

\section{ Results and Discussions} \label{sec_sim}

\subsection{Simulation Setup}
We consider a dense pixel-based FAS at the reader with sub-wavelength inter-port spacing
\begin{math}
d = \lambda/8,
\end{math}
which enables a compact antenna aperture and strong spatial correlation among ports.
The carrier frequency is set to $f_\mathrm{c}=3.5$~GHz, corresponding to a wavelength $\lambda \approx 0.0857$~m.
The FAS consists of $K \in \{5, 10, 20\}$ uniformly spaced ports over a linear aperture of length $W\lambda$, where $W=(K-1)d/\lambda$.
The RF source continuously transmits an unmodulated carrier with a maximum transmit power $P_\mathrm{s}=20$dbm. The noise power at the reader is set to $\sigma^2=-100$~dBm and
the circuit power consumption of the BD is $P_\mathrm{c}=-20$~dBm.
For imperfect observation modeling, the variance of the relative channel gain observation error is set to $\sigma_\delta^2=0.05$, corresponding to a moderate level of measurement uncertainty in the observed cascaded channel gain.
Large-scale path loss between the RF source, the BD, and the reader follows
\begin{math}
L(d)=d^{-\chi},
\end{math}
where $\chi=2.9$ denotes the path-loss exponent.
The forward (source-to-B)D and backscatter (BD-to-FAS) channels incorporate both large-scale attenuation and small-scale fading,
which is modeled as spatially correlated Rician fading with the backscatter channel $\mathbf{h}_\mathrm{b}=[h_{\mathrm{b},1},\dots,h_{\mathrm{b},K}]^\mathsf{T}$ is generated as
\begin{equation}
\mathbf{h}_\mathrm{b} = \sqrt{\frac{\kappa}{\kappa+1}}\mathbf{h}_\mathrm{LOS}
+ \sqrt{\frac{1}{\kappa+1}}\mathbf{R}^{1/2}\mathbf{w},
\end{equation}
where $\kappa=5$ is the Rician factor, $\mathbf{w}\sim\mathcal{CN}(\mathbf{0},\mathbf{I})$,
and $\mathbf{R}$ is the spatial correlation matrix constructed using the Jakes' model in \eqref{eq:jakes}.
The matrix square root $\mathbf{R}^{1/2}$ is obtained via Cholesky factorization.
The forward channel $h_\mathrm{s}$ is modeled independently with the same Rician factor.

The OFF state in BD's OOK modulation corresponds to $\alpha(0)\approx0$, while the ON state is characterized by
\begin{math}
\alpha(1)=\sqrt{a_1},
\end{math}
where $a_1\in[a_{\min},a_{\max}]$ is optimized subject to the energy-neutrality constraint.
Unless otherwise stated, the bit probability is set to $p_1=\Pr(b=1)=0.5$.
The reflection efficiency is fixed to $\eta=0.8$, and the RF-to-DC conversion efficiency is set to $\xi=0.6$.
All performance metrics are computed using the OTA approach by averaging over $N=10^5$ independent channel and observation-noise realizations.
The PSO algorithm employs a swarm of $N_p=50$ particles and runs for $T=50$ iterations.
The inertia weight is set to $\omega=0.6$, and the cognitive and social learning coefficients are chosen as $c_1=c_2=1.2$.
As a benchmark, we consider a traditional antenna system (TAS) reader equipped with a single antenna, where perfect CSI in a single-input single-output (SISO) AmBC system is assumed to be provided at the reader.

\subsection{Result Analysis and Discussion}
%%%%%%%%%%%%%%%%%%%%%%%%%%%%
\begin{figure}[t]
    \centering    \includegraphics[width=0.4\textwidth]{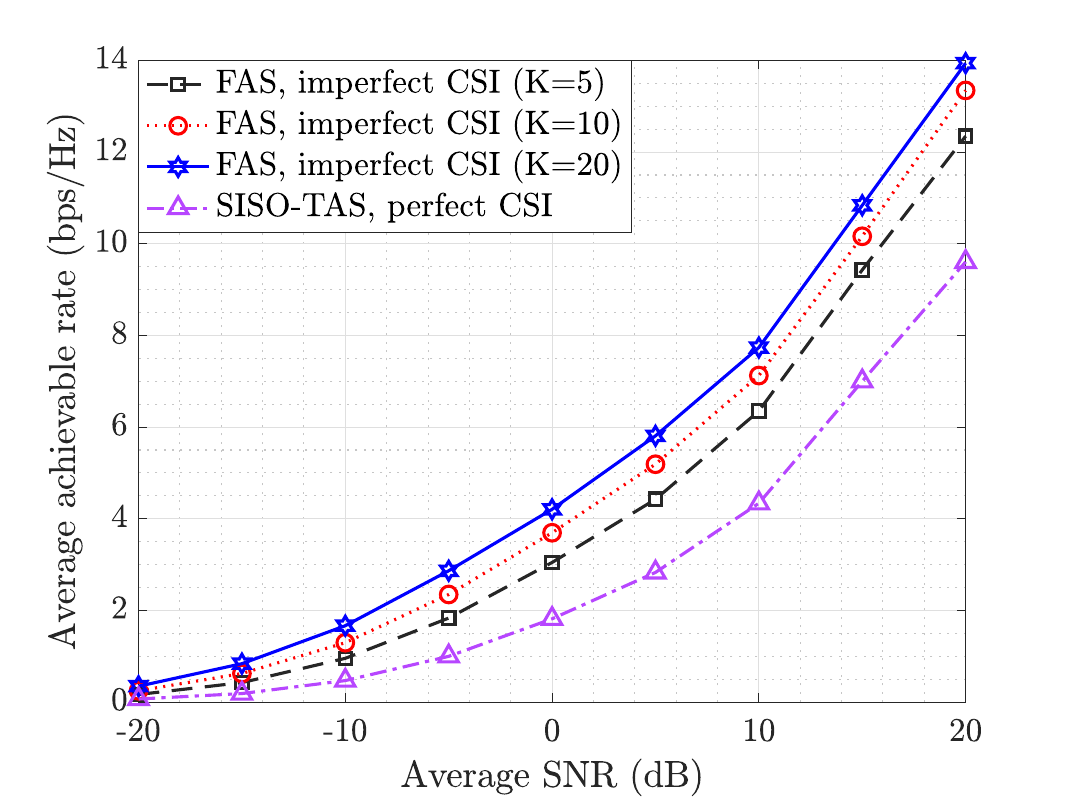}
    \caption{Average achievable rate versus average SNR for the proposed FAS-based AmBC with different number of preset positions.}
    \label{fig:Rate_SNR_K}
\end{figure}
%%%%%%%%%%%%%%%%%%%%%%%%%%%%
Fig.~\ref{fig:Rate_SNR_K} plots the average achievable rate versus the average received SNR and highlights the clear advantage of the proposed pixel-based FAS under imperfect CSI relative to the SISO-TAS benchmark even with perfect CSI. Across the full SNR range, the FAS curves remain above TAS, showing that the spatial selectivity gained by switching among multiple ports provides a diversity benefit that outweighs occasional port-misselection under noisy channel observations. The performance improves steadily with the number of candidate ports, with \(K\!=\!20\) outperforming \(K\!=\!10\) and \(K\!=\!5\) at all SNRs, because a larger port set increases the chance of encountering a favorable cascaded-channel realization and thus raises the effective selected gain. At low SNR, the curves are relatively close since noise dominates and selection diversity has limited impact; at moderate and high SNR, the achievable rate becomes more sensitive to the selected cascaded gain inside \(\log_2(1+\gamma)\), so the gap between FAS and TAS, as well as the separation among different \(K\), becomes more visible. 
%These results confirm that pixel-FAS offers a robust and scalable means of exploiting spatial diversity in AmBC links, delivering consistent rate gains under realistic imperfect-CSI operation.

%%%%%%%%%%%%%%%%%%%%%%%%%%%%
\begin{figure}[t]
    \centering    \includegraphics[width=0.4\textwidth]{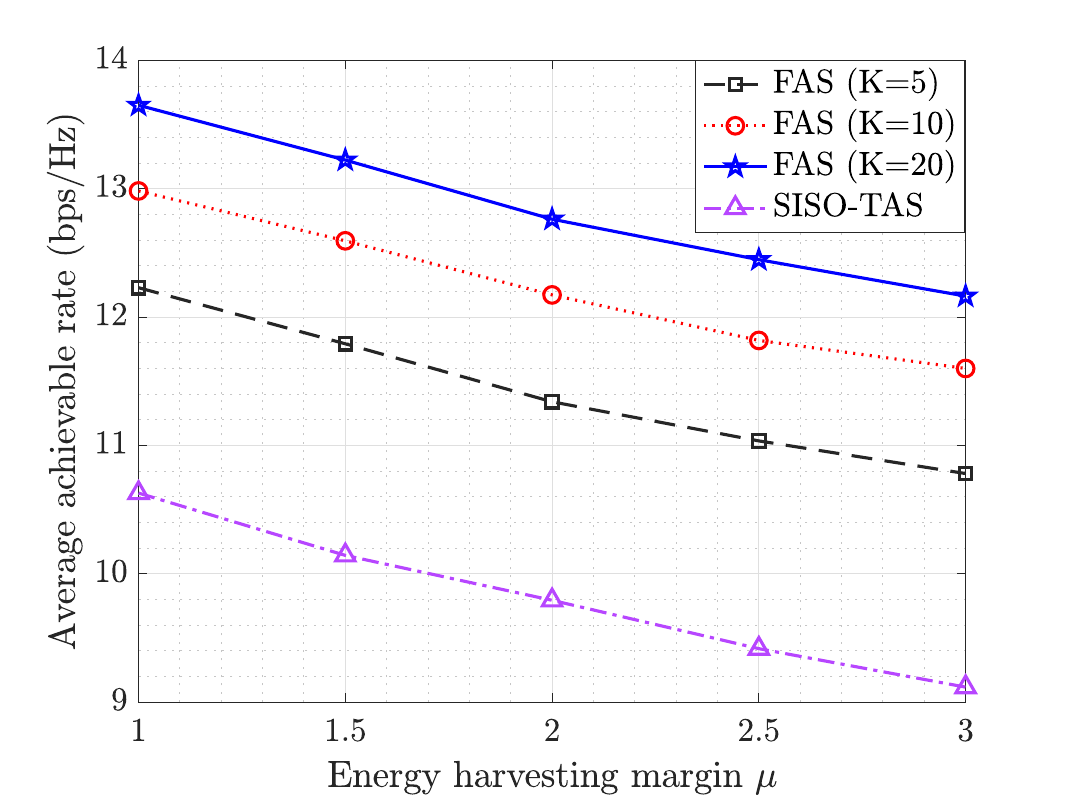}
    \caption{Average achievable rate versus average SNR for the proposed FAS-based AmBC with different number of preset positions.}
    \label{fig:Rate_EH_margin}
\end{figure}
%%%%%%%%%%%%%%%%%%%%%%%%%%%%
Fig.~\ref{fig:Rate_EH_margin} illustrates the impact of the EH safety margin $\mu$ on the average achievable rate for the proposed pixel-based FAS AmBC. As $\mu$ increases, the harvested energy requirement at BD becomes more stringent, which directly tightens the feasible reflection coefficient and reduces the portion of incident power available for backscatter modulation. This effect leads to a monotonic degradation in achievable rate for all schemes. Despite this increasingly conservative energy constraint, the pixel-FAS consistently maintains a clear performance advantage over the TAS benchmark across the entire range of $\mu$, highlighting the robustness of spatial diversity even when energy availability is constrained. Moreover, increasing the number of FAS pixels yields a systematic rate improvement, as larger $K$ provides a higher probability of selecting a favorable backscatter path under imperfect CSI. The parallel downward trends observed for all configurations confirm that the safety margin primarily governs the energy-rate tradeoff, while the relative ordering of the curves demonstrates that the gains from pixel-level spatial selection remain effective under strict EH requirements.

%%%%%%%%%%%%%%%%%%%%%%%%%%%%
\begin{figure}[t]
    \centering    \includegraphics[width=0.4\textwidth]{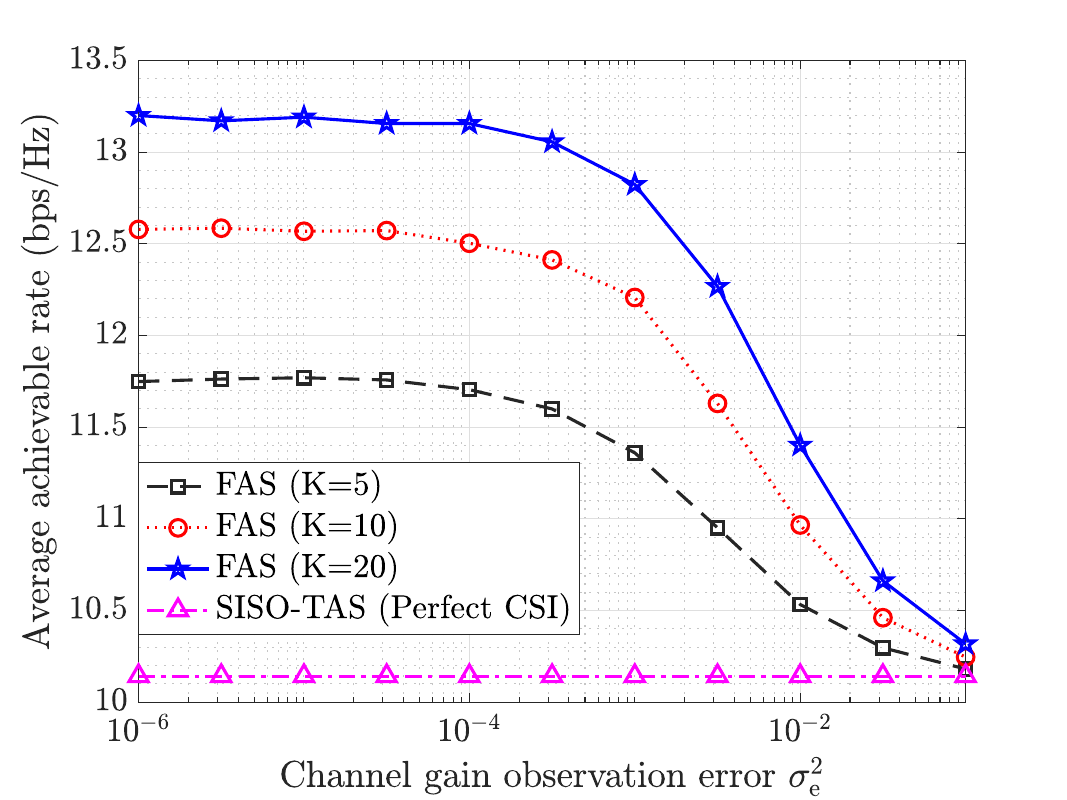}
    \caption{Average achievable rate versus CSI error variance for the proposed FAS-based AmBC with different number of preset positions.}
    \label{fig:Rate_CSI}
\end{figure}
%%%%%%%%%%%%%%%%%%%%%%%%%%%%
Fig.~\ref{fig:Rate_CSI} illustrates the impact of channel gain observation error variance on the average achievable rate. For very small error variance, the achievable rate of all FAS configurations remains nearly constant, indicating that pixel selection based on the observed cascaded channel gains is reliable when the gain observation is sufficiently accurate. As the observation error variance increases beyond approximately \(10^{-3}\), a clear degradation in achievable rate is observed for all FAS cases. This degradation becomes more pronounced as the number of pixels \(K\) increases, since a larger candidate set amplifies the probability of incorrect pixel selection when the gain observations are noisy. 
In contrast, the SISO-TAS benchmark exhibits almost invariant performance over the entire range of error variance, as no antenna or pixel selection is involved and the received signal depends only on the true channel realization. Despite its sensitivity to gain observation uncertainty, the pixel-based FAS scheme consistently outperforms the SISO-TAS benchmark across a wide range of error levels. This confirms that the spatial diversity offered by multiple pixels provides a substantial rate advantage that outweighs the adverse effects of imperfect gain observation, except in regimes where the observation noise becomes severely dominant.

%%%%%%%%%%%%%%%%%%%%%%%%%%%%
\begin{figure}[t]
    \centering    \includegraphics[width=0.4\textwidth]{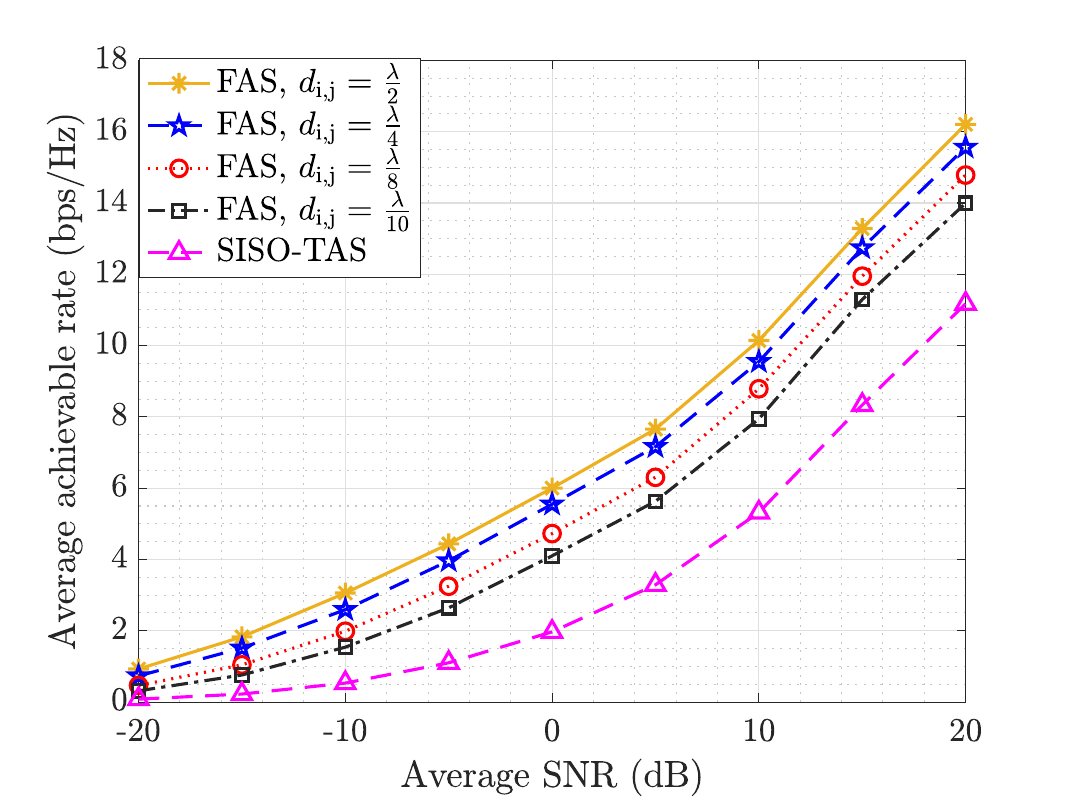}
    \caption{Average achievable rate versus average received SNR for the pixel-based FAS-aided AmBC with $K=10$ and different inter-pixel spacings $d_\text{i,j}$.}
    \label{fig:Rate_lambda}
\end{figure}
%%%%%%%%%%%%%%%%%%%%%%%%%%%%
Fig.~\ref{fig:Rate_lambda} shows the average achievable rate versus SNR for the proposed pixel-based FAS under different inter-pixel spacings $d_{i,j}$. Several trends are evident. First, for all SNR values, the FAS consistently outperforms the SISO-TAS reader, highlighting the spatial diversity and aperture gains of multi-pixel structures. Second, increasing the inter-pixel spacing improves the achievable rate, with $d_{i,j}=\lambda/2$ yielding the best performance due to reduced spatial correlation and an effectively larger aperture, which increases the likelihood of selecting a strong cascaded channel.
At moderate and high SNR, the performance gap between spacings becomes more pronounced, emphasizing the role of spatial separation when noise is less dominant. However, increasing the spacing does not yield unbounded gains. Beyond a certain point, the rate saturates as the benefits of reduced correlation diminish and path-loss, finite aperture, and channel hardening effects dominate. This suggests the existence of an optimal spacing that balances spatial diversity and practical constraints.

\section{Conclusion} \label{sec_conclusion}
This paper investigated a pixel-based FAS architecture for AmBC receivers under practical EH and channel uncertainty constraints. By exploiting antenna position flexibility within a compact aperture, the proposed FAS-enabled reader mitigates the severe effects of cascaded double path loss and multiplicative fading without increasing RF chains or compromising the low-power nature of AmBC. 
The joint optimization of FAS port selection and backscatter modulation was formulated as a mixed discrete–continuous problem under an EH constraint. To address its nonconvexity and imperfect channel observations, a PSO-based framework was developed to obtain near-optimal solutions. The OTA metric was adopted to capture practical operation based on noisy observations.
Simulation results demonstrated that the proposed FAS-enabled AmBC consistently outperforms a conventional single-antenna reader across diverse settings, while remaining robust to observation errors and strict energy constraints. The results also highlight the tradeoff among spatial diversity, energy availability, and aperture utilization, showing that performance gains saturate for a fixed physical surface.

\section*{Acknowledgment}
{This work has received funding from the SNS JU under the EU’s Horizon Europe research and innovation programme under Grant Agreement No. 101192113 (Ambient-6G), the EU's Horizon 2022 Research and Innovation Programme under Marie Skłodowska-Curie Grant
No. 101107993, and by grant PID2023-149975OB-I00 (COSTUME) funded by MICIU/AEI/10.13039/501100011033 and by FEDER/UE.
}

\end{document}